\documentclass[twocolumn]{webofc}

\usepackage[varg]{txfonts}   
\begin{document}
\title{GdFe$_2$ Laves phase intermetallic system under pressure: an ab-initio study}
%
%

\author{\firstname{Elena} \lastname{Kokorina}\inst{1}\fnsep\thanks{\email{kokorina@iep.uran.ru}} \and
        \firstname{Michael} \lastname{Medvedev}\inst{1}\fnsep\thanks{\email{medvedev@iep.uran.ru}}
\and
\firstname{Igor} \lastname{Nekrasov}\inst{1}\fnsep\thanks{\email{nekrasov@iep.uran.ru}}}

\institute{Institute of Electrophysics UB RAS, 620016, Ekaterinburg, Russia}

\abstract{%
Here we perform $ab-initio$ study of Curie temperature $T_C$ under hydrostatic pressure for intermetallic compound GdFe$_2$.
To calculate $T_C$ for GdFe$_2$ we applied mean-field solution of the Heisenberg model for several magnetic sublattices with
DFT/LDA calculated values of necessary exchange interaction integrals and local magnetic moments.
To compare with available experimental data pressure values were taken from zero up to about 70 Kbar. It corresponds to
2\% compression of the volume of the unit cell. In agreement with experimental data $T_C$ grows under pressure.
It was shown that Fe ions magnetic sublattice alone provides only about 75\% of the experimental Curie temperature $T_C^{exp}$.
Gd sublattice is found to give very weak contribution to the $T_C^{exp}$. Here we show that the missing 25 \% of $T_C^{exp}$ comes from
Fe-Gd exchange pairs.
}
\maketitle
\section{Introduction}

Intermetallic compounds of rare-earth elements with transition metals are extremely diverse in their magnetic properties and promising for practical applications \cite{Wallace1973,Buschow_1991}. Among them, ferrimagnetic compounds of gadolinium with iron are of considerable interest, since strong exchange interactions in the iron atom subsystem provide relatively high Curie temperatures $T_C$, and the large magnetic moments of gadolinium atoms in some cases give a large magnetic moment per intermetallic formula unit.

Note that in the crystallographic series of Gd-Fe intermetallic compounds there are compounds of the type Gd$_2$Fe$_{17}$ with the maximum iron content in the formula chemical unit, and GdFe$_2$ with the minimum iron content. The magnitude of the magnetic moments and exchange interactions in the iron sublattice in Gd$_2$Fe$_{17}$ for the two crystallographic phases were calculated both at zero pressure $P=0$ \cite{Lukoyanov2009} and at finite hydrostatic pressure \cite{Igoshev2018}.

In this case, the widespread hypothesis about the existence of competition between ferromagnetic and antiferromagnetic exchange interactions between the nearest magnetic neighbors in the iron atom subsystem was disproved. In this paper, we study the behavior of magnetic moments and exchange interactions at zero pressure $P$=0 and hydrostatic compression at the other end of a series of Gd-Fe intermetallic compounds, namely for GdFe$_2$.

The intermetallic compound GdFe$_2$ has a crystal structure of the MgCu2 type (cubic symmetry C15), and the unit cell of a stoichiometric compound contains 2 formula units in a volume $V=a_0^3$ \cite{Tesluk1969}, where the lattice parameter $a_0=7.4\AA $ is taken at zero pressure $P=0$ \cite{Creagh1978}. The nearest neighbors of the Gd atom are 12 Fe ions and 4 Gd ions; the nearest neighbors of the Fe atom are 6 Gd ions and 6 Fe ions \cite{Tesluk1969}.

The electronic structure of GdFe$_2$ is calculated by the LSDA+U method \cite{Anisimov1997}, within the framework of the atomic sphere approximation in the basis of the linearized MT-orbitals (TB-LMTO-ASA v. 47)\cite{LMTO1,LMTO2,LMTO3}. Coulomb interactions for Gd-4f states were taken as $U$=6.7 eV and $J$=0.7 eV, for the Fe-3d shell - $U$=2 eV and $J$=0.9 eV. Necessary Heisenberg exchange integral values are obtained as described in Ref. \cite{Anisimov1997}. The first Brillouin zone was splitted in k-space as 8x8x8. Previously, the choice of orbital basis and atomic spheres radii for Fe and Gd were justified in Ref. \cite{Lukoyanov2009}. This particular setup was used here also for non zero pressure $P\neq$0 calculations.

\section{Results an discussion}

To calculate the change in the magnetic moments and the exchange interactions in GdFe$_2$ with pressure, the equilibrium crystal lattice parameter $a$ needs to be matched to the corresponding hydrostatic pressure value $P$. For this, the Murnaghan model \cite{Murnaghan1944,Palasyuk2004} is used, in which the relationship between pressure $P$ and the relative change in the volume of the unit cell $V_0/V(P)$ is given by the following state equation
\begin{equation}
P=\frac{B_0}{B^\prime_0}\left[\left({\frac{V_0}{V(P)}}\right)^{B^\prime_0} - 1\right],
\end{equation}
where $B_0$ -- bulk compression module and $B^\prime_0$ -- its first derivative with respect to pressure. For GdFe$_2$ these values are
 $B_0$=1040.5 Kbar and $B^\prime_0$=4.49 \cite{Gomaa2015}. The correspondence of the  lattice parameters relative changes $(a-a_0)/a_0$ and pressures $P$ is given in Table 1.

The calculation of the change in magnetic moments under pressure in GdFe$_2$ shows that as the pressure increases, the resulting magnetic moment of the formula unit increases smoothly (see Table 1). This is due to the fact that in the pressure range under study, the magnetic moments of 2 Fe atoms, which are antiparallel to the moment of the Gd atom, decrease under compression, whereas the magnetic moment of the Gd atom remains almost unchanged. The reasons for this are obvious - the 3d electron shells in the Fe atom are significantly more distant from the core than the 4f shells in Gd, and therefore they deform more strongly under pressure than the 4f shells of a rare-earth metal.

As for the direct exchange interaction between the iron atoms in GdFe$_2$, it has a ferromagnetic character, is quite large and increases with decreasing interatomic distance under all-round compression (Table 1). At the same time, even at zero pressure, it is comparable in magnitude with the exchange parameters between Fe atoms in the dumbbell positions of the Gd$_2$Fe$_{17}$ intermetallic compound \cite{Lukoyanov2009} (for GdFe$_2$ $I_{Fe-Fe}$=257.4 K at interatomic distance $r_{Fe-Fe}$=2.616\AA, whereas in Gd$_2$Fe$_{17}$ $I_{Fe1-Fe1}$=238.8 K at $r_{Fe-Fe}$=2.4\AA~for hexagonal structure and for rhombohedral structure of Gd$_2$Fe$_{17}$ $I_{Fe1-Fe1}$=287.5 K at $r_{Fe-Fe}$=2.385\AA).

The obtained exchange value $I_{Fe-Fe}(P)$ (see Table 1) makes it possible to estimate the Curie temperature $T_C$ in the molecular field approximation for classical spin vectors the nearest neighbors interactions inside the iron sublattice by the formula
\begin{equation}
k_B T_C = \frac{S^2_{Fe}}{3} I_{Fe-Fe}z_{Fe-Fe},
\end{equation}
here $z_{Fe-Fe}$=6 -- number of Fe nearest neighbors gr  a given Fe site and $S_{Fe}=m_{Fe}/g\mu_B=m_{Fe}/2\mu_B$ -- classical spin vector value. At zero pressure $P$=0 corresponding $T_C$(Fe)=600.5 K. That value is considerably lower than experimental value $T^{exp}_C$=802.5 K \cite{Brouha1973}. If such estimates are made at a non zero pressure $P\neq$0, then the Curie temperature $T_C$ increases with increasing pressure P. But all the time it lies about 200 K below the experimental values of the experimental $T_C$ under pressure. It is obvious that realistic estimates of the $T_C$ in GdFe$_2$ are impossible without taking additional account of the interactions between the subsystems of the Fe and Gd atoms.

We first note that within the framework of the LSDA+U approach, the parameters of the direct exchange between Gd atoms $I_{Gd-Gd}$ were also calculated. The exchange parameter $I_{Gd-Gd}$ turned out to be antiferromagnetic, varying slightly with a change of pressure and 3 orders of magnitude smaller in absolute value than the ferromagnetic exchange parameter $I_{Fe-Fe}$. Obviously, a very weak antiferromagnetic exchange $I_{Gd-Gd} < 0$ cannot have a noticeable effect on the formation of the high Curie temperature $T_C$ in GdFe$_2$.

As for the exchange parameter between the ions Gd and Fe $I_{Gd-Fe}$, at present, within the framework of band calculations, a computational procedure has not been developed for finding the parameters of the direct exchange between atoms whose partially filled electronic shells have different orbital moments, for example, between the 3d shell atoms and the atoms with 4f shell. Therefore, for an approximate estimate of the magnitude of the exchange $I_{Gd-Fe}$ in GdFe$_2$, we use the indirect method.

When generalized to the case of classical spins, the Curie temperature of a two-sublattice magnetic system in the approximations of the molecular field and the interaction of the nearest neighbors has the form \cite{Neel1948,Brouers1971}
\begin{eqnarray}
k_B T_C &=& \frac{1}{6}\{S^2_{A}I_{AA}z_{AA} + S^2_{B}I_{BB}z_{BB}+\\ \nonumber
        &+&[(S^2_{A}I_{AA}z_{AA} - S^2_{B}I_{BB}z_{BB})^2+ \\ \nonumber
        &+&4 S^2_{A}S^2_{B}I^2_{AB}z_{AB}z_{BA}]^{\frac{1}{2}}\},
\label{eq3}
\end{eqnarray}
where A means Fe, B means Gd. Note that, in the general case, the numbers of the nearest neighbors in different atom pairs are not equal to each other $z_{AB}\neq z_{BA}$. Thus in GdFe$_2$ corresponding numbers are $z_{Fe-Gd}$=6 but $z_{Gd-Fe}$=12 \cite{Tesluk1969}. It follows from Eq.(\ref{eq3}) that
\begin{equation}
I^2_{AB}=\frac{(3k_B T_C - S^2_{A}I_{AA}z_{AA})(3k_B T_C - S^2_{B}I_{BB}z_{BB})}{S^2_{A}S^2_{B}z_{AB}z_{BA}}.
\end{equation}

Using the $I_{Fe-Fe}$ and $I_{Gd-Gd}$ values calculated by the LSDA+U method and the experimental values of the Curie temperature $T_C^{exp}$ \cite{Brouha1973}, we estimated the exchange parameter $I_{Fe-Gd}$ at different pressures. It turned out (see the bottom line in Table 1) that the parameter $I_{Fe-Gd}$ is negative and provide antiferromagnetic exchange. The absolute value of$I_{Fe-Gd}$ is about 6 times less than one of the ferromagnetic exchange $I_{Fe-Fe}$. Also $I_{Fe-Gd}$ value slightly increases under compression. However, it should be borne in mind that the results for the $I_{Fe-Gd}$ exchange value obtained using the molecular field approximation and $T_C^{exp}$ are valid under assumption that GdFe$_2$ is a Heisenberg magnet.

\section{Summary}

$Ab-initio$ study of Curie temperature $T_C$ under hydrostatic pressure for intermetallic compound GdFe$_2$ is performed.
Corresponding Curie temperatures $T_C$ for GdFe$_2$ are obtained within mean-field solution of the Heisenberg model for several magnetic sublattices.
The required values of Heisenberg exchange interaction integrals and local magnetic moments were calculated based on DFT/LDA results.
Here we used all-round compression of  GdFe$_2$ unit cell down to 2\% of its volume which corresponds to the pressure up to about 70 Kbar.
In agreement with experimental data theoretically calculated $T_C$ grows under pressure.
However Fe ions magnetic sublattice along provides only about 75\% of the experimental Curie temperature $T_C^{exp}$.
On the other hand Gd sublattice gives very weak contribution to the $T_C^{exp}$. Finally, in this work, we have shown that another 25\% of $T_C^{exp} $ are provided due to exchange interactions in the Fe-Gd pair.

\section{Acknowledgments}
The work was carried out within the theme of state contract N0389-2015-0024 and with partial support of the RFBR project N18-02-00281.

\begin{table*}
\centering
\caption{Relative lattice parameter change, corresponding lattice parameters and pressure; LSDA+U calculated pressure dependence of different GdFe$_2$ parameters: local magnetic moments for different sites, values of Heisenberg exchange interaction for different pairs Fe-Fe and Gd-Fe; experimental Curie temperature $T^{exp}_C$ and corresponding values of Heisenberg exchange interaction for Fe-Gd pairs.}
\label{tab-1}       
\begin{tabular}{llllllll}
\hline
$\frac{a-a_0}{a_0}$ &    0   & -0.0025 & -0.005 & -0.0075 & -0.01  & -0.015 & -0.02   \\\hline
$a$, \AA            & 7.400  & 7.382   & 7.363  & 7.344   & 7.326  & 7.289  & 7.252   \\\hline
P, Kbar             & 0      & 7.95    & 16.19  & 24.73   & 33.60  & 52.32  & 72.48   \\\hline
$m_{Fe},~\mu_B$     & -2.16  & -2.15   & -2.14  & -2.13   & -2.12  & -2.10  & -2.08   \\\hline
$m_{Gd},~\mu_B$     &  6.94  &  6.94   &  6.94  &  6.94   &  6.94  &  6.94  &  6.93   \\\hline
$m_{GdFe_2},~\mu_B$ &  2.62  &  2.64   &  2.66  &  2.68   &  2.70  &  2.74  &  2.77   \\\hline
$I_{Fe-Fe}$, K      & 257.42 & 261.33  & 262.91 & 275.08  & 278.79 & 286.85 &  293.08 \\\hline
$I_{Gd-Gd}$, K      & -0.19  & -0.19   & -0.19  & -0.19   & -0.19  & -0.21  &  -0.21  \\\hline
$T^{exp}_C$, K      & 802.5  & 805.0   & 807.7  & 810.4   & 813.2  & 819.2  & 825.7   \\\hline
$I_{Fe-Gd}$, K      & -38.05 & -38.19  & -38.89 & -37.25  & -37.53 & -38.03 &  -39.07 \\\hline
\end{tabular}
\end{table*}
%
%
%

\end{document}